\documentclass[12pt]{article}
\usepackage{epsfig}
\def\be{\begin{eqnarray}}
\def\ee{\end{eqnarray}}
\begin{document}
\title{Nuclear matter EOS with a three-body force}
\author{ \normalsize A. Lejeune$^{1}$, U. Lombardo$^{2}$, W. Zuo$^{3}$\\
\normalsize 
$^{1}$ Institut de Physique, B5 Sart-Tilman, B-4000 Li\`ege 1, 
Belgium \\ \normalsize
$^{2}$ Dipartimento di Fisica, 57 Corso Italia, I-95129 Catania, Italy \\
\normalsize 
and INFN-LNS, 44 via Santa Sofia, I-95123 Catania, Italy \\ \normalsize
$^{3}$ Institute of Modern Physics, Lanzhou, China }
\maketitle
\begin{abstract}
The effect of a microscopic three-body force on the saturation properties of
nuclear matter is studied within the Brueckner-Hartree-Fock approach. 
The calculations show a decisive improvement of the saturation density along
with an overall agreement with the empirical saturation point.
With the three-body force the symmetry energy turns more rapidly increasing 
with density, which allows for the direct URCA process to occur in   
$\beta$-stable neutron star matter. The influence of the three-body force
on the nuclear mean field does not diminish the role of the ground state 
correlations. 

\end{abstract} 

\newpage
\section{ Introduction }

The theoretical prediction of the nuclear equation of state (EOS) 
is of great interest for understanding extreme states of matter as to 
density, temperature and isospin. In spite of the great deal of effort
already done, yet this topic is largely controversial in many aspects, 
including the mechanism of nuclear saturation.

In non-relativistic approaches the model of nucleons interacting only 
via a two-body force fails to reproduce the empirical saturation 
observables. Thus phenomenological three-body forces (TBF), derived 
by the Urbana group, have been introduced with few adjustable parameters 
within both the variational method \cite{PAND,WIR,AKM1,AKM2} and 
the Brueckner theory \cite{BAL,BURG}. 

At the contrary, the success of Dirac-Brueckner calculations without 
TBF \cite{MALF,MACH,LEE} seems to indicate that the mechanism of 
saturation is a purely relativistic effect. This effect can be traced 
to the virtual excitations of nucleon-antinucleon pairs and their 
contribution alone, estimated in Ref.~\cite{BROW}, seems to be 
able to reproduce the correct saturation properties \cite{TAOR}. 

This raises the question of the role played by other important 
elementary processes such as the nucleonic 
excitations ($\Delta$ or Roper resonance). This point was examined 
already a decade ago by Grange et al. \cite{LEJ}, who proposed a 
microscopic TBF based on the meson-exchange model. In that paper
they reported a decisive improvement of both the saturation energy and 
density with respect to the existing Brueckner-Hartree-Fock (BHF) 
predictions based on pure two-body interactions \cite{CUGN}. Later 
more refined BHF calculations with two-body
force gave results so sharply different \cite{REVI} to call for a 
numerical re-examination of the BHF approach with the microscopic TBF.

In the present note this re-examination is made and new results for 
the EOS of nuclear matter are reported which benefit from a 
more sophisticated version of the two-body realistic interaction, 
i.e., the Argonne $AV18$ \cite{WIRI}. The calculations are 
also extended to neutron matter with the purpose of extracting the 
symmetry energy which over the last years has been raising a strong 
interest for its remarkable astrophysical implications \cite{ZUO}.     

\section{Effective three-body force}

Our microscopic three-body force is deduced from the meson-exchange 
current approach. It contains the contribution due to the medium 
modification of the 
two-meson ($\pi\pi,\pi\rho,\rho\rho$) exchange part of the 
nucleon-nucleon ($NN$) interaction, the contribution associated to 
the $\sigma$ and $\omega$ meson exchange and, finally, the 
$\rho\pi\gamma$ diagram. Their bare mass and free coupling constant 
are assigned to all mesons except the $\sigma$ meson, for which a 
strong mass renormalization is expected as medium polarization 
effect. Moreover its value cannot be completely dissociated from 
the other parameters giving the two-body force due to the 
self-consistent nature of the present approach \cite{MATH}. 
The value of $540$MeV has been adopted with the Paris 
force \cite{LEJ}, but it is also compatible with the $AV18$, which 
is used in the present calculations.   
For a more detailed description of the model and approximations 
we refer to Ref.~\cite{LEJ}.  
 
The effect of the TBF has been included in the calculation along 
the same line as in~\cite{LEJ}, where it is conveniently reduced 
to an effective two-body interaction to avoid the difficulty of the 
full three-body problem. A detailed description and justification 
of the procedure is given in \cite{LEJ}. Here we simply write down 
the equivalent two-body potential, which is given in $r$-space by
\be
 \langle\vec r_1 \vec r_2| V_3 | \vec{r}_1^{\ \prime}  
 \vec{r}_2^{\ \prime} \rangle =
 \frac{1}{4} Tr\sum_n \int {\rm d}{\vec r_3} 
{\rm d} {\vec{r}_3^{\ \prime}} 
\phi^*_n(\vec r_3{\ \prime} ) [1-\eta(r_{13}' )] 
[1-\eta(r_{23}')] 
\nonumber
\ee
\be
\times 
W_3(\vec r_1^{\ \prime} \vec r_2^{\ \prime} 
\vec r_3^{\ \prime}|\vec r_1 \vec r_2 \vec r_3)
\phi_n(r_3) 
[1-\eta(r_{13})] [1-\eta(r_{23})]
\ee  
where the trace is taken with respect to the spin and isospin of 
the nucleon 3. The function $\eta(r)$ is the average over spin and 
momenta in the Fermi sea of the defect function, in which only the 
most important partial wave components have been included, i.e., 
the $^1S_0$ and $^3S_1$ partial waves. According to Eq.~(1) the 
effective two-body interaction is obtained by averaging the 
three-body force over the wave function of the third particle
taking into account the correlations between this particle and the 
two others.

As above mentioned the mass of $\sigma$ meson is taken as a free 
parameter in the present TBF. This is due to the lack of information 
on the medium polarization effect on the $\sigma$ meson propagator. 
The EOS could offer a chance of investigating this effect as far as 
the two-body force is also built on a meson-exchange model. 
For a numerical comparison with the results Ref.~\cite{LEJ}, 
we decided to adopt the same value of the $\sigma$-meson mass, 
i.e., $m_{\sigma}=540~MeV$, 
which anyhow makes the saturating effect of the TBF to be the most 
efficient.

\section { Numerical results }

Due to its dependence on the defect function the effective two-body 
interaction, Eq.~(1), is calculated selfconsistently at each step of 
the iterative BHF procedure. Otherwise the procedure is the same as 
in standard BHF approach. The $G$-matrix is calculated selfconsistently 
along with the auxiliary potential by solving the Bethe-Goldstone 
equation. The continuous choice is adopted for the auxiliary potential 
because it provides a convergence of the hole-line expansion much 
faster than the gap choice \cite{SONG}. The Argonne $AV18$ is used 
for the two body nuclear force as it gives an excellent fit to $NN$ 
scattering data as well as to the deuteron binding energy \cite{WIRI}.  

\subsection{EOS in nuclear and neutron matter} 

In Fig.~1 the energy per nucleon is reported for both nuclear matter 
(lower curves) and neutron matter (upper curves). The effect of the 
three-body force can be assessed by the comparison between the BHF 
results with (solid curve) and without (short dashed curve) 
three-body force. The three-body force affects the EOS in such 
a way that in the low-density domain the energy per nucleon is 
practically unchanged, but in the high-density domain it rises 
up significantly. One main result for the EOS of symmetric nuclear 
matter is that the saturation density turns decisively improved 
towards the empirical value, from $0.265 fm^{-3}$ to $0.19 fm^{-3}$. 
At the same time there is no appreciable change in the saturation 
energy. This is a quite desirable feature of the three-body force, 
since the result for the saturation energy in our BHF calculation 
based on purely two-body interaction is already in good agreement 
with its empirical counterpart. All that makes us more confident 
with the general guess on the importance of the three-body force 
for the saturation problem of nuclear force. Compared to the old 
calculation with Paris force in Ref.~\cite{LEJ} the 
new saturation density obtained in the present work is 
smaller and closer to the empirical value. This is not 
a surprise because in the calculation of Ref.~\cite{LEJ} the 
saturation density predicted with only a two-body force was a bit 
larger than $0.3 fm^{-3}$. 
\vspace*{1cm} 
\begin{figure}[ht]
\centerline{\epsfig{figure = 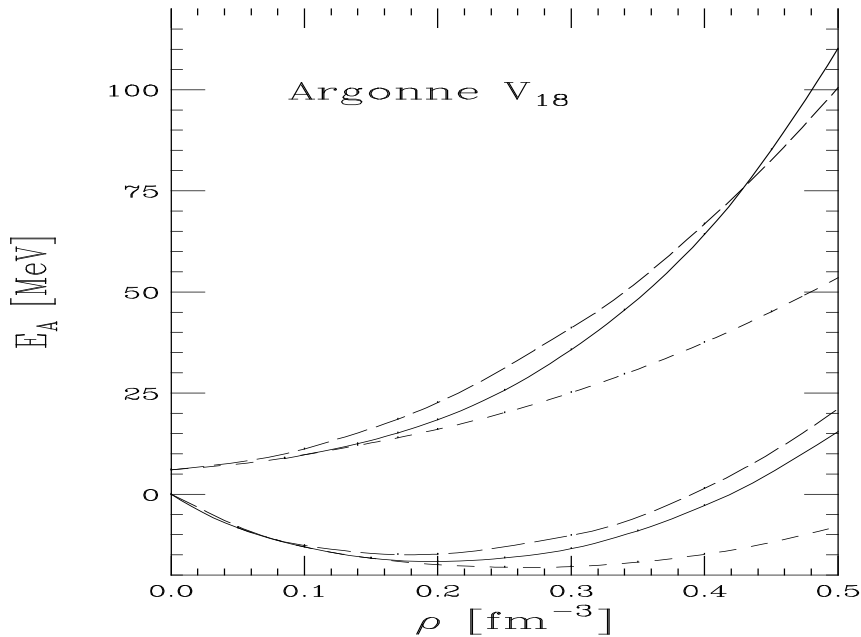,angle=90,width=12cm,height=8cm}}
\caption{\small EOS of nuclear matter (lower curves) and neutron 
matter (upper curves). The Brueckner calculations are 
plotted for the $AV18$ without TBF (short-dashed lines) and 
with the TBF (solid lines) discussed in the text. For 
comparison are also reported the calculations with 
AV18 and Urbana TBF of Ref.~[6] (long-dashed lines). }
\end{figure}\normalsize

In Fig.~1 is also plotted the EOS's obtained from BHF calculations 
with two-body $AV18$ force and phenomenological Urbana 
TBF \cite{BURG} (longdashes). Despite the two different TBF's 
the two EOS's do not significantly differ from each other both 
for nuclear matter and for neutron matter. The Urbana TBF has two 
free parameters which were adjusted \cite{BAL} in order to reproduce 
the saturation point, as shown in the latter EOS. 

In recent variational calculations, which adopt the $AV18$ combined 
with the recent Urbana $IX$ TBF \cite{AKM1,AKM2}, the two free 
parameters of the TBF are fixed to reproduce the binding energy 
of $^3 H$ and the saturation density of nuclear matter. But the 
resulting equilibrium energy of about $-12 MeV$ for symmetric 
nuclear matter disagrees with the empirical value and no significant 
improvement is achieved by including relativistic boost 
corrections \cite{AKM2}. This discrepancy casts some doubt on 
constraining the TBF by means of the binding energy of light nuclei.

No less important is the effect on the stiffness of the EOS 
measured by the compressibility modulus
\be
     K \,\,=\,\, 9 \rho^2 {\partial^2 E_A \over \partial \rho^2} 
\ee
calculated at the saturation density. The effect of three-body force 
on the compressibility is to enhance its value from $250$MeV to 
$265$MeV. This small change is due to the fact that the more 
pronounced curvature of the saturation curve is balanced by the 
smaller value of the saturation density.   

The neutron matter EOS combined with that of symmetric nuclear 
matter provides us information on the isospin effects \cite{ZUO}, 
in particular on the symmetry energy.
The symmetry energy is defined as
\be
E_{sym}(\rho) \quad = \quad {1\over 2} \left[ {\partial^2 E_A(\rho,\beta)
\over \partial \beta ^2}\right]_{\beta=0}.
\ee
It is well established \cite{BOMB} that the binding energy 
per nucleon $E_A$ fulfills the simple $\beta^2$-law not only for 
$\beta \ll 1$ as assumed in the empirical nuclear mass 
formula \cite{MASS}, but also in the whole asymmetry range. This 
enables us to calculate the symmetry energy $E_{sym}$ in terms of 
the difference between the binding energy of pure 
neutron matter $E_A(\rho,1)$ and that of symmetric nuclear matter 
$E_A(\rho,0)$, i.e., 
\be
E_{sym}(\rho) \,=\, E_A(\rho,1) - E_A(\rho,0), 
\ee
but one would refrain from applying it at very high density. 
The results of our calculation for the symmetry energy as a 
function of baryonic density are depicted in Fig.~2. 
\begin{figure}[ht]
\centerline{\epsfig{figure = 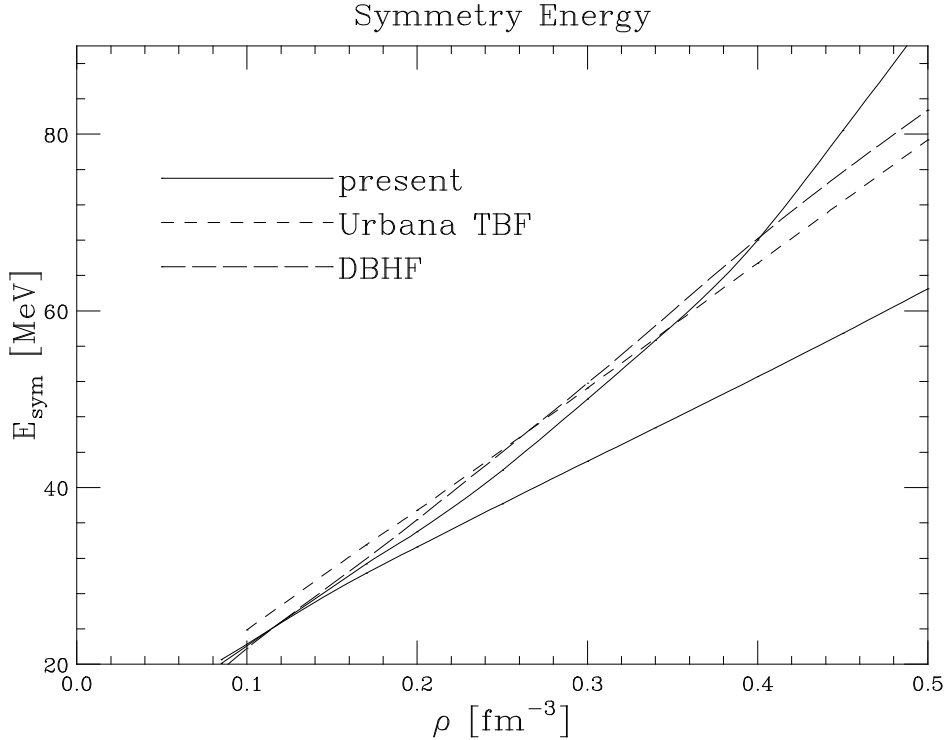,angle=90,height=10cm}}
\caption{\small Density dependence of the symmetry energy in different 
approaches. The two solid lines correspond to the present 
approach without TBF (lower curve) and with TBF (upper curve), 
the short dashed curve to the EOS's with Urbana TBF (Ref.~[6]) and the 
long dashed curve to a DBHF calculation with the Bonn 
one-boson-exchange potential (Ref.~[9]). }
\end{figure}\normalsize

As expected from the strongly repulsive TBF component at high 
density, a considerable enhancement of $E_{sym}$ is 
found (compare the two solid lines). At the saturation point 
the enhancement is rather small, i.e., from 30.3 to 31.3MeV, 
which anyway is in agreement with the empirical value of 
$30 \pm 4$MeV \cite{MASS}. Also reported are the symmetry energy 
obtained from the Urbana TBF \cite{BURG} and the DBHF 
calculation \cite{LEE}. Apart from the density range above 
0.4fm$^{-3}$, all three curves exhibit an almost linear increase 
of $E_{sym}$($\rho$) as a function of $\rho$ and also a quantitative 
agreement with one another. A rather good agreement with the 
variational approach is also found extracting the symmetry energy 
from the EOS's reported in Refs.~\cite{AKM1,AKM2}. The difference, 
which for instance at $\rho = 0.4$fm$^{-3}$ is a bit less than $15\%$, 
would be attributed to the fact that the variational calculation 
underbinds symmetric nuclear matter.     

The close similarity of our result with the DBHF prediction  
is quite astonishing if one considers the two 
different contexts, i.e. the relativistic DBHF without a TBF 
in Ref.~\cite{LEE} and the present non-relativistic BHF with TBF, 
in which they have been obtained. On the other hand, we already 
mentioned the equivalence between the medium effects on the Dirac 
spinor in the Dirac-Brueckner approach and the virtual excitation 
of $N\bar N$ pairs, which is an important component of the present 
TBF \cite{TAOR}. In spite of the good agreement in $E_{sym}$, 
one expects significant differences for observables which are 
very sensitive to $E_{sym}$ such as the proton fraction 
$Y = \frac{Z}{A}$ in $\beta$-equilibrium nuclear matter, 
which approximately fulfills a power law , i.e., 
$Y \simeq E_{sym}^3$ \cite{PAIR}. Due to the important 
implications for the cooling mechanism in neutron stars \cite{LATT}, 
a comparison among the different predictions is worth to be done. 
The proton fraction is calculated from the equilibrium condition 
in nuclear matter formed by mixture of neutrons, protons, electrons 
and muons with respect to weak interaction. Assuming charge 
neutrality, one easily gets
\be
3\pi^2 (\hbar c)^3 \rho Y  \,\,=\,\, 
[4E_{sym}(1-2Y)]^3 + 
\nonumber
\ee
\be
\{ [4E_{sym}(1-2Y)]^2 - m_{\mu}c^2\}^{3/2}
\theta(\mu_e-m_{\mu}c^2)
\ee 
where the step function $\theta(\mu_e-m_{\mu}c^2)$ indicates that 
when the electron chemical potential exceeds the muon rest mass, 
the muon channel opens and muons participate to the chemical 
equilibrium. Within this context we calculated the proton fractions 
corresponding to the different $E_{sym}$ reported in Fig.~2. 
The values of $Y(\rho)$ are reported in Tab.~I. 
It has been shown \cite{LATT} that the direct URCA process can 
occur if the proton fraction exceeds certain threshold values of $Y$, 
which has been estimated of about 0.148 in the case 
of $\mu_e \gg m_{\mu}c^2$. This value 
serves as a good approximation for a simple estimate. 
For example, at the density $\rho=0.45$, the critical 
value of $Y$ is about 0.136 for electrons and about 0.161 for muons.  
It is seen that in the density domain here 
considered (up to 0.5fm$^{-3}$) this value is not reached in the 
non-relativistic calculation based on purely two-body force, 
whereas it is reached in the range 0.4 -- 0.5 fm$^{-3}$ in the other 
cases. Comparing the data of last column 
with the data of Ref.~\cite{BURG}, where the $AV14$ is used as two-body 
potential, we notice that the $AV18$ interaction has the effect of 
reducing the density threshold from 0.5fm$^{-3}$ to a value 
between 0.4 and 0.5fm$^{-3}$. 
\begin{center}
\small{Tab.~I  \ \ Proton fractions vs density for $\beta$-equilibrium 
nuclear matter: our results from Argonne $V18$ without (second column) 
and with TBF (third column), Dirac-Brueckner calculations from 
Ref.~[9] (forth column) and and BHF from $AV18$+Urbana from 
Ref.~[6] (last column). }
\vskip 0.2cm
\begin{tabular}{ccccc}
\hline
 $\rho$ & BHF & BHF+TBF & DBHF & BHF+Urbana \\ 
 \hline
 0.04 &   0.0161 &   0.0138  &           &         \\                   
 0.06 &   0.0210 &   0.0181  &           &         \\ 
 0.085&   0.0245 &   0.0229  &           &         \\
 0.10 &   0.0264 &   0.0264  &   0.0241  &   0.031 \\
 0.14 &   0.0328 &   0.0338  &           &         \\
 0.17 &   0.0378 &   0.0417  &   0.0403  &   0.050 \\
 0.20 &   0.0440 &   0.0508  &   0.0528  &   0.060 \\
 0.30 &   0.0659 &   0.0920  &   0.0993  &   0.096 \\
 0.40 &   0.0865 &   0.1365  &   0.1364  &   0.128 \\
 0.50 &   0.1069 &   0.1882  &   0.1616  &   0.155 \\
\hline
\end{tabular}
\end{center}\normalsize
\vspace*{.5cm}

Before we conclude, let us give a brief discussion about the 
single-particle ( s.p. ) potential. 
It is found that at low density ($\rho\le\rho_0=0.17$fm$^{-3}$), 
the effect of the TBF on the s.p. potential is rather small. 
At the saturation density $\rho_o=0.17$fm$^{-3}$, only a small 
increase about $1.0 \sim 1.5$MeV in symmetric nuclear matter 
is observed at the momentum around the Fermi momentum $k_F$, 
which amounts
a reduction of the attractive mean field of less that $2 \%$. 
This is consistent with the result for the binding energy $E_A$ 
as discussed before and suggests that it is still necessary, even 
after introducing the TBF, to include the ground state 
particle-hole exitations (the second-order $M_2$ contributions) in 
the mass operator in order to get a satisfactory agreement 
to the phenomenological optical potential \cite{ZUO,GRANG}. 
As the density increases, the repulsive contribution of the TBF to 
the s.p. potential becomes increasingly pronounced since the TBF 
plays its major role at high density. For example, 
at $\rho=0.34$fm$^{-3}$ for symmetric nuclear matter, there is an 
enhancement over the whole momentum range (from $-$130.8 to 
$-$123.7MeV at $k=0$ and $-$94.4 to $-$80.3MeV at $k_F$). 
Again this is in agreement with the observation on the binding 
energy $E_A$. 

\section{Summary and conclusions} 

In summary, we have investigated the effect of a microscopic 
three-body force on the EOS of symmetric nuclear matter and pure 
neutron matter within the Brueckner approach.
The introduction of a TBF turns out to be crucial for reproducing 
the empirical saturation point, confirming the previous investigations 
with phenomenological TBF. But the microscopic TBF has the advantage 
of tracing the properties of nuclear matter back to more fundamental 
degrees of freedom. In addition it is suitable for a closer contact 
with relativistic approaches. 

In particular, the saturation density approaches closely the empirical 
value after including the TBF in the calculation. 
At the contrary the saturation energy turns out to be almost unaffected 
which is a desirable result since, in the present calculation, 
its value only with two-body force is already in good agreement 
with the empirical binding energy. Accordingly, 
the TBF modifies appreciably the mean field only above the saturation 
density, so confirming the limit of BHF approximation without 
ground state correlations in describing the phenomenological optical 
potential. From the calculation of the EOS for pure neutron matter we 
have extracted the symmetry energy, which turns in good agreement 
with the relativistic DBHF calculations and also with the BHF and 
variational calculations adopting phenomenological Urbana TBF. 

The steep uprise of the energy symmetry with density could 
have deep influence on the properties of neutron stars and on 
the supernovae explosions. The proton fraction has been calculated 
under conditions of $\beta$-equilibrium. The threshold value necessary 
to switch on the direct URCA processes is reached below 0.5fm$^{-3}$ 
when the three-body force is included. This result may have remarkable 
consequences for the neutron-star cooling mechanism. 

Finally, we address the question whether or not the EOS of nuclear 
matter can probe the medium modifications of the $NN$ interaction. 
Our approach to the TBF looks suitable for such a purpose provided 
that the two-body interaction is also described in terms of a 
meson-exchange model.
    
\section*{Acknowledgments}
We wish to thank J.F. Mathiot for valuable discussions and G.F. Burgio 
for providing us the results with $AV18$ and Urbana TBF.


\begin{thebibliography}{99}
\bibitem{PAND}
B. Friedman and V. R. Pandharipande, Nucl. Phys. {\bf A361}, 502 (1981).
\bibitem{WIR} 
R. B. Wiringa, V. Fiks, and A. Fabrocini, Phys. Rev. {\bf C38}, 
 1010 (1988).
\bibitem{AKM1} A. Akmal and V. R. Pandharipande, Phys. Rev.
{\bf C56}, 2261 (1997). 
\bibitem{AKM2} A. Akmal, V. R. Pandharipande and D. G. Ravenhall, Phys. Rev.
{\bf C58}, 1804 (1998). 
\bibitem{BAL}
 M. Baldo, I. Bombaci, and G. F. Burgio, Astron. and Astrophys. {\bf 328}, 274 
 (1997).
\bibitem{BURG} F. Burgio, private communication.
\bibitem{MALF} 
B. ter Haar and R. Malfliet, Phys. Reports {\bf 149}, 208 (1987).
\bibitem{MACH} 
R. Machleidt, Adv. Nucl. Phys. {\bf 19}, 189 (1989) and 
references therein quoted.
\bibitem{LEE} 
C.-H. Lee, T. T. S. Kuo, G. Q. Li and G. E. Brown, Phys. Rev. {\bf C57},
 3488 (1998).
\bibitem{BROW} 
G. E. Brown, W. Weise, G. Baym and J. Speth, Comm. Nucl. Part.
Phys. {\bf 17}, 39 (1987).
\bibitem{TAOR} 
M. Baldo, G. Giansiracusa, U. Lombardo, I. Bombaci and L. S. Ferreira, 
Proceedings of the Vth International Conference on "Nucleus-Nucleus
Collisions", Taormina (Italy) 1994. M. Di Toro, E. Migneco and 
P. Piattelli Eds. Nucl. Phys. {\bf A583}, 599 (1995).
\bibitem{LEJ} 
P. Grang\'e, A. Lejeune, M. Martzolff, and J.-F. Mathiot, 
Phys. Rev. {\bf C40}, 1040 (1989). 
\bibitem{CUGN} 
A. Lejeune, P. 
Grang\'e, M. Martzolff and J. Cugnon, Nucl. Phys.
{\bf A453}, 189 (1986).
\bibitem{REVI}
H.-J. Schulze, J. Cugnon, A. Lejeune, M. Baldo, 
and U. Lombardo, Phys. Rev. {\bf C52}, 2785 (1995).
\bibitem{WIRI} 
R. B. Wiringa, V. G. J. Stoks, and R. Schiavilla, 
Phys. Rev. {\bf C51}, 38 (1995).
\bibitem{ZUO} 
W. Zuo, I. Bombaci and U. Lombardo, Phys. Rev. {\bf C60}, 024605 (1999) 
and references therein quoted.                  
\bibitem{MATH} J.-F. Mathiot, private communication.
\bibitem{SONG}
H. Q. Song, M. Baldo, G. Giansiracusa, and U. Lombardo, 
Phys. Rev. Lett. {\bf 81}, 1584 (1998)
\bibitem{BOMB} 
I. Bombaci and U. Lombardo, Phys. Rev. {\bf C44}, 1892 (1991).
\bibitem{MASS} 
P. E. Haustein, Atomic Data and Nuclear Data Tables {\bf 39}, 185 (1988).
\bibitem{PAIR} M. Baldo, J. Cugnon, A. Lejeune and U. Lombardo,
Nucl. Phys. {\bf A536}, 349(1991).
\bibitem{LATT}
J. Lattimer, C. Pethick, M. Prakash and P. Haensel, Phys. Rev. Lett. 
{\bf 66}, 2701 (1991).
\bibitem{GRANG} 
P. Grang\'e, J. Cugnon and A. Lejeune, Nucl. Phys. {\bf A473}, 365 (1987).
\end{thebibliography}
\end{document}